%% file: main.tex
\documentclass[conference]{IEEEtran}
\IEEEoverridecommandlockouts
\usepackage{paralist}
\usepackage{cite}
\usepackage{amsmath,amssymb,amsfonts}
\usepackage{algorithmic}
\usepackage{graphicx}
\usepackage{textcomp}
\usepackage{xcolor}
\usepackage{acronym}
\usepackage{booktabs}
\usepackage{enumitem}
\usepackage{tikz}
\usetikzlibrary{positioning,calc,shapes.geometric}
\usepackage{url}
\def\BibTeX{{\rm B\kern-.05em{\sc i\kern-.025em b}\kern-.08em
    T\kern-.1667em\lower.7ex\hbox{E}\kern-.125emX}}

\DeclareMathOperator*{\argmin}{arg\,min}
\begin{document}

\newcommand{\tcpu}{T_{\rm cpu}^{\rm FL}}
\newcommand{\tfl}{T_{\rm tx}^{\rm FL}}
\newcommand{\tra}{T_{\rm tx}^{\rm RA}}
\newcommand{\tidle}{T_{\rm idle}^{\rm FL}}
\newcommand{\sfl}{S^{\rm FL}}
\newcommand{\sra}{S^{\rm RA}}
\def\figW{10}

\newcommand{\flicon}{figures/graphics-card.png}
\newcommand{\raicon}{figures/motion-sensor.png}
\newcommand{\bsicon}{figures/antenna.png}

\tikzset{
  fldevice/.style={
    shape=rectangle,
    inner sep=0pt,
    outer sep=0pt,
    minimum width=5mm,
    minimum height=5mm,
    draw=none,
    path picture={
      \node at (path picture bounding box.center)
        {\includegraphics[width=5mm]{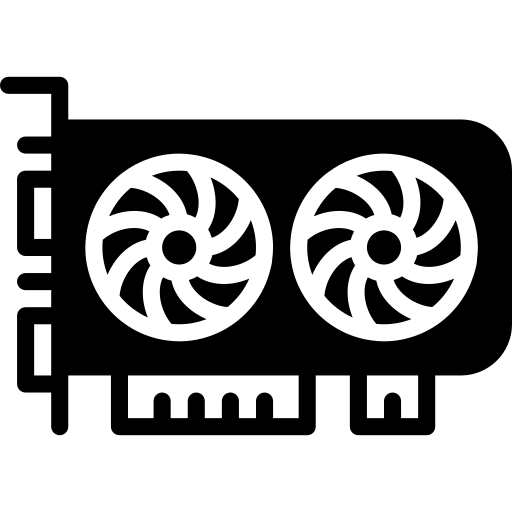}};
    }
  },
  radevice/.style={
    shape=rectangle,
    inner sep=0pt,
    outer sep=0pt,
    minimum width=5mm,
    minimum height=5mm,
    draw=none,
    path picture={
      \node at (path picture bounding box.center)
        {\includegraphics[width=5mm]{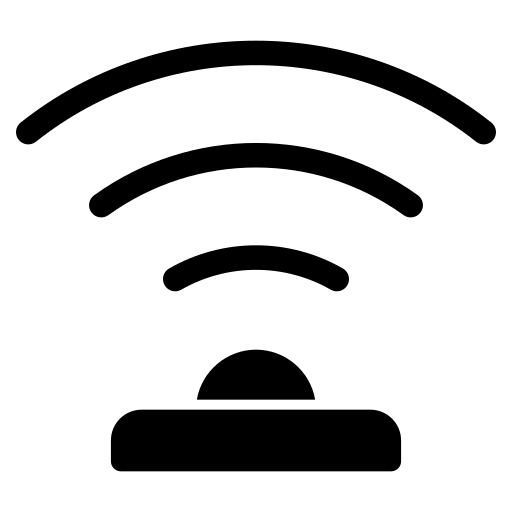}};
    }
  },
  bsnode/.style={
    shape=rectangle,
    inner sep=0pt,
    outer sep=0pt,
    minimum width=9mm,
    minimum height=9mm,
    draw=none,
    path picture={
      \node at (path picture bounding box.center)
        {\includegraphics[width=9mm]{\bsicon}};
    }
  }
}

\acrodef{eMBB}{enhanced mobile broadband}
\acrodef{FDMA}{frequency division multiple access}
\acrodef{FL}{federated learning}
\acrodef{GenAI}{generative artificial intelligence} 
\acrodef{IoT}{internet of things}
\acrodef{MAC}{medium access control}
\acrodef{ML}{machine learning}
\acrodef{RAN}{radio access network}
\acrodef{saloha}[S-ALOHA]{slotted-ALOHA}
\acrodef{SGD}{stochastic gradient descent}
\acrodef{URLLC}{ultra-reliable low latency communication}
\acrodef{RA}{random access}
\acrodef{AI}{artificial intelligence}

\newcommand\copyrighttext{%
  \footnotesize This work has been submitted to the IEEE for possible publication. Copyright may be transferred without notice, after which this version may no longer be accessible.}
\newcommand\copyrightnotice{%
\begin{tikzpicture}[remember picture,overlay]
\node[anchor=south,yshift=10pt] at (current page.south) {\fbox{\parbox{\dimexpr\textwidth-\fboxsep-\fboxrule\relax}{\copyrighttext}}};
\end{tikzpicture}%
}

\newcommand{\tred}[1]{\textcolor{red}{#1}}
\newcommand{\tblue}[1]{\textcolor{blue}{#1}}
\newcommand{\cache}[1]{} 

\title{Federated Learning Meets Random Access: Energy-Efficient Uplink Resource Allocation 
\thanks{This work has been supported by the EU through the Horizon Europe/JU SNS project ROBUST-6G (grant no. 101139068).}
}

\author{\IEEEauthorblockN{Giovanni Perin \thanks{
G. Perin is also with the Dept. of Information Engineering (DEI), University of Padova, Padova, Italy. 
}}
\IEEEauthorblockA{\textit{Dept. of Inform. Eng. (DII)} \\
\textit{University of Brescia}\\
Brescia, Italy \\
giovanni.perin@unibs.it}
\and
\IEEEauthorblockN{Eunjeong Jeong}
\IEEEauthorblockA{\textit{Dept. of Comp. and Inform. Sci.} \\
\textit{Link\"oping University}\\
Link\"oping, Sweden \\
eunjeong.jeong@liu.se}
\and
\IEEEauthorblockN{Nikolaos Pappas}
\IEEEauthorblockA{\textit{Dept. of Comp. and Inform. Sci.} \\
\textit{Link\"oping University}\\
Link\"oping, Sweden \\
nikolaos.pappas@liu.se}
}

\maketitle

\copyrightnotice

\begin{abstract}
Artificial intelligence-generated traffic is changing the shape of wireless networks. Specifically, as the amount of data generated to train machine learning models is massive, network resources must be carefully allocated to continue supporting standard applications. In this paper, we tackle the problem of allocating radio resources for two sets of concurrent devices communicating in uplink with a gateway over the same bandwidth. A set of devices performs federated learning (FL), and accesses the medium in FDMA, uploading periodically large models. The other set is throughput-oriented and accesses the medium via random access (RA), either with ALOHA or slotted-ALOHA protocols. We derive close-to-optimal solutions to the non-convex problem of minimizing the system energy consumption subject to FL latency and RA throughput constraints. Our solutions show that ALOHA can sustain high FL efficiency, yielding up to 48\% lower consumption when the system is dominated by FL traffic. On the other hand, slotted-ALOHA becomes more efficient when RA traffic dominates, yielding 6\% lower consumption.
\end{abstract}

\input{sections/intro}

\input{sections/model}

\input{sections/formulation}

\input{sections/results}

\input{sections/concl}

\bibliographystyle{IEEEtran}
\bibliography{IEEEabrv, biblio}

\end{document}

%% file: sections/intro.tex
\section{Introduction}

Wireless networks have traditionally carried traffic with predictable, downlink-heavy patterns. However, the steep rise of \ac{AI} applications is changing wireless traffic characteristics~\cite{mahajan2025:GenAI-nokia}. Resource scarcity has arisen as a key challenge again, where efficient allocation between ``AI tasks'' and ``communication tasks'' is critical~\cite{news:nvidia}. Notably, unlike conventional mobile traffic that follows approximately a 90:10 downlink-to-uplink ratio, \ac{GenAI} applications exhibit a substantially higher uplink share, which reaches 26\% of total traffic~\cite{GenAI-traffic:ericsson}. This introduction of bandwidth-intensive new media formats will affect future mobile network traffic volumes and characteristics, particularly through changing uplink requirements~\cite{GenAI-mobilityreport:ericsson}.

Resource allocation for heterogeneous service coexistence has been extensively studied, including \ac{FL} resource allocation~\cite{dinh2021:FL-resource-allocation, yang2021energy, chen2021:joint-FL-WN, ballotta2025vremfl} and the coexistence of edge inference and \ac{FL}~\cite{luo2024efficient, han2024federated}. Among related studies,  joint optimization of sensing (gathering data), computation (inference), and communication (\ac{FL} updates) \cite{liang2025joint, xu2025:ILAC} have covered wireless network systems consisting of clients with different traffic profiles, often represented as \ac{URLLC} and \ac{eMBB} users~\cite{ganjalizadeh2025, adhikari2024, alsenwi2021:URLLC-eMBB, anand2020:URLLC-eMBB}. However, throughput-oriented traffic, exemplified by emerging \ac{GenAI} inference applications, presents fundamentally different challenges~(C) from the existing frameworks.
\begin{enumerate}[label=\textbf{(C\arabic*)}, leftmargin=*, align=left]
    \item\textbf{Different traffic characteristics.} \ac{URLLC}/\ac{eMBB} literature addresses the negotiation problem between services prioritizing high data rate versus high reliability and low latency~\cite{adhikari2024}. \ac{URLLC} traffic consists of small, sparse packets with strict latency and reliability requirements, making latency as the primary challenge~\cite{popovski2019:URLLC}. In contrast, throughput-oriented traffic such as IoT sensor updates~\cite{centanaro2026longrange} and \ac{GenAI} inference streams~\cite{zhang2025:beyond_the_cloud} often involves frequent transmissions with throughput guarantees as the dominant requirement. The conflict thus shifts from latency interference to bandwidth exhaustion: Clients focusing on distributed model training and AI inference compete for limited spectrum resources.
    \item\textbf{Lack of relevant works despite significance and timeliness.} While several works have explored integrating \ac{GenAI} \emph{into} \ac{FL}~\cite{ning2024:fedGCS, guo2024:promptFL, li2024filling} or \ac{FL} \emph{into} \ac{GenAI}~\cite{chen2024exploring, huang2024federated}, few studies consider scenarios where \ac{FL} and throughput-oriented services operate as \emph{parallel, competing} frameworks sharing wireless channel bandwidth. A recent work on joint \ac{FL} and inference optimization~\cite{li2021:joint-AI-training-inference} addresses resource-constrained edge devices but assumes homogeneous traffic types and does not account for the distinct medium access mechanisms.
\end{enumerate}
 
To address this gap, we propose a joint radio resource allocation framework for two client groups: \ac{FL} devices (aiming model convergence) and throughput-oriented devices (requiring guaranteed throughput). In this paper, we formulate the bandwidth allocation problem as the minimization of the global communication energy consumption subject to an FL latency budget and a minimum throughput guarantee for random access devices. We derive close-to-optimal solutions by jointly optimizing the bandwidth share and the transmission rate, and compare two \ac{MAC} protocols for the throughput-oriented devices. Numerical experiments support the optimization results and corroborate the strategy to choose MAC protocols based on given conditions. 

%% file: sections/model.tex
\section{System Model and Analysis}
\label{sec:system}
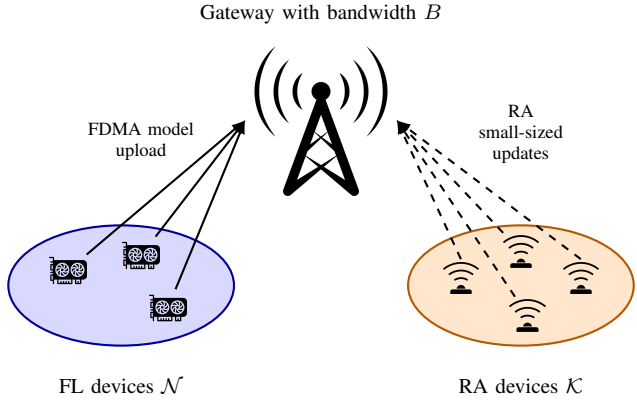
\begin{figure}[t]
\centering
\begin{tikzpicture}[
    x=\columnwidth/\figW,
    y=1cm,
    >=latex,
    node distance=1.0cm and 2.0cm,
    flgroup/.style={
  ellipse,
  draw=blue!60!black,
  thick,
  fill=blue!15
    },
    ragroup/.style={
      ellipse,
      draw=orange!70!black,
      thick,
      fill=orange!20
    }
]

\node[inner sep=0] (bs) at (0,0)
  {\includegraphics[width=2cm]{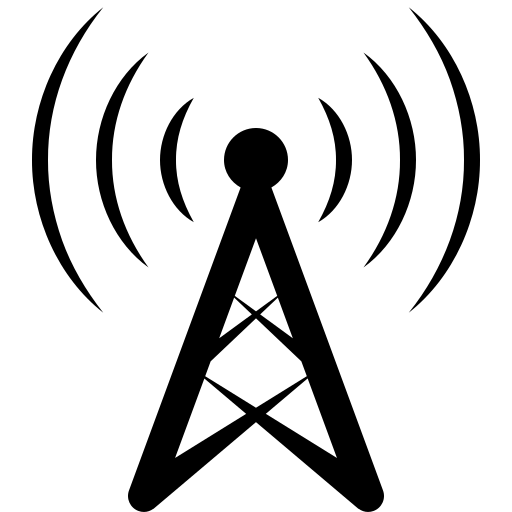}};

\node[flgroup, minimum width=3cm, minimum height=1.6cm] (flcluster) at (-3,-2.2) {};
\node[below=0.3cm of flcluster] {\footnotesize FL devices $\mathcal{N}$};

\node[fldevice] (fl1) at ($(flcluster.center)+(0.3,0.4)$) {};
\node[fldevice] (fl2) at ($(flcluster.center)+(0.7,-0.3)$) {};
\node[fldevice] (fl3) at ($(flcluster.center)+(-0.8,0.2)$) {};

\node[ragroup, minimum width=3cm, minimum height=1.6cm] (racluster) at (3,-2.2) {};
\node[below=0.3cm of racluster] {\footnotesize RA devices $\mathcal{K}$};

\node[radevice] (ra1) at ($(racluster.center)+(-0.9,0.1)$) {};
\node[radevice] (ra2) at ($(racluster.center)+(0,0.45)$) {};
\node[radevice] (ra3) at ($(racluster.center)+(0.9,0.1)$) {};
\node[radevice] (ra4) at ($(racluster.center)+(0.1,-0.4)$) {};

\draw[->, thick] (fl1) -- (bs.west);
\draw[->, thick] (fl2) -- (bs.west);
\draw[->, thick] (fl3) -- (bs.west);

\draw[->, thick, dashed] (ra1.north) -- (bs.east);
\draw[->, thick, dashed] (ra2) -- (bs.east);
\draw[->, thick, dashed] (ra3.north) -- (bs.east);
\draw[->, thick, dashed] (ra4) -- (bs.east);

\node[font=\footnotesize, above=0.15cm of bs] {Gateway with bandwidth $B$};

\node[font=\scriptsize, align=center, text width=2cm, above=0.8cm of ra2]
  {\ac{RA}\\small-sized\\updates};

\node[font=\scriptsize, align=center, text width=2cm, above=0.9cm of fl1]
{FDMA model\\upload};

\end{tikzpicture}
    \caption{System model. A gateway serves two classes of devices sharing the same uplink channel: FL devices transmit model updates in FDMA and throughput-oriented devices acces the channel via \ac{RA} (ALOHA / S-ALOHA).}
    \label{fig:system-model}
    \vspace{-0.3cm}
\end{figure}

We consider a wireless network where a gateway with available bandwidth $B$ offers communication services to two disjoint sets of nodes differing in the task they perform. Specifically, set $\mathcal{N}=\{1,\ldots,N\}$ denotes the nodes participating in the federated training of a \ac{ML} model, using a \ac{FDMA} protocol at the \ac{MAC} layer; the set $\mathcal{K}=\{1,\ldots,K\}$ denotes instead the nodes producing throughput-oriented traffic, e.g., \ac{GenAI} inference streams characterized by frequent and bandwidth-intensive uplink transmissions. We consider the latter devices use a \ac{RA} protocol at the \ac{MAC} (ALOHA or \ac{saloha}). The system is shown in Fig.~\ref{fig:system-model}.

\subsection{Federated learning communication model}
\begin{figure}[tb]
\centering
\begin{tikzpicture}
  \def\xzero{0}
  \def\xone{2.6}
  \def\xtwo{3.6}
  \def\xthree{5.4}

  \definecolor{set2orange}{RGB}{252,141,98}
  \definecolor{set2blue}{RGB}{141,160,203}
  \definecolor{set2green}{RGB}{166,216,84}

  \draw[->] (\xzero-0.5,0) -- (\xthree+0.5,0) node[right] {$t$};

  \foreach \x/\t in {
    \xzero/t_0,
    \xone/t_1,
    \xtwo/t_2,
    \xthree/t_3
  }{
    \draw (\x,0.1) -- (\x,-0.1) node[below] {$\t$};
  }

  \draw[line width=2pt,set2orange] (\xzero,0) -- (\xone,0);
  \draw[line width=2pt,set2blue]   (\xone,0) -- (\xtwo,0);
  \draw[line width=2pt,set2green]  (\xtwo,0) -- (\xthree,0);

  \node[anchor=south] at ({(\xzero+\xone)/2},0.1)
    {$\overbrace{\hspace{2.3cm}}^{\tcpu}$};
  \node[anchor=south] at ({(\xone+\xtwo)/2},0.1)
    {$\overbrace{\hspace{0.9cm}}^{\tidle}$};
  \node[anchor=south] at ({(\xtwo+\xthree)/2},0.1)
    {$\overbrace{\hspace{1.6cm}}^{\tfl}$};
\end{tikzpicture}
\caption{Scheme of the \ac{FL} round. Instants $t_0, t_1, t_2$ and $t_3$ denote the beginning/end of different phases, while $\tcpu, \tidle$, and $\tfl$ refer to the intervals of computation, idling, and transmission, respectively.}
\label{fig:FL-timings}
\vspace{-0.3cm}
\end{figure}
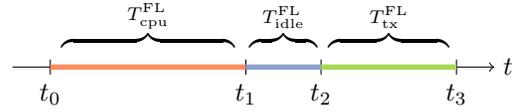

Devices belonging to set $\mathcal{N}$ alternate computation (local \ac{SGD} steps) and transmission (local model upload), after possibly waiting for some idle time $\tidle \ge 0$. The computation time is denoted by $\tcpu$, and is considered a parameter; on the other hand, the transmission time, denoted by $\tfl$, is a variable quantity. As a consequence, each learning round has a duration
\begin{equation}
    \label{eq:learning-duration}
    T_{\rm tot}^{\rm FL} = \tcpu + \tidle + \tfl.
\end{equation}
We denote as $t_0,t_1,t_2$, and $t_3$ the beginning instants of computation, idling, transmission, and the round end instant, respectively, as shown in Fig.~\ref{fig:FL-timings} (it is possible that $t_1 \equiv t_2$). During time $\tfl$, device $n \in \mathcal{N}$ updates the gateway in \ac{FDMA}, transmitting with power $P_{\rm tx}^{\rm FL}$ at Shannon rate
\begin{equation}
    \label{eq:fl-rate}
    R_n^{\rm FL}(\rho) = \rho \frac{B}{N} \log_2\left(1+\frac{g_n P_{\rm tx}^{\rm FL}}{\rho \frac{B}{N} N_0}\right),
\end{equation}
where $g_n$ is the channel gain between device $n$ and the gateway, $N_0$ is the noise power, and $\rho \in [0,1]$ is the bandwidth share allocated to \ac{FL} devices.

Hence, denoting by $\sfl$ the \ac{ML} model size, the global \ac{FL} transmission time can be written as
\begin{equation}
\begin{aligned}
    \label{eq:fl-tx-time}
    \tfl(\rho) &= \max_n\left\{T_n^{\rm FL}(\rho)\right\} = \frac{\sfl}{\min_n\left\{R_n^{\rm FL}(\rho)\right\}}=\\
    &=\frac{N\sfl}{\rho B \log_2\left(1+\frac{g_n^{\rm min} N P_{\rm tx}^{\rm FL}}{\rho B N_0}\right)},
\end{aligned}
\end{equation}
where $T_n^{\rm FL}$ denote the individual transmission times and $g_n^{\rm min} = \min_n\{g_n\}$. The transmission energy per device is thus
\begin{equation}
    \label{eq:transmission-energy}
    E_n^{\rm FL}(\rho) = P_{\rm tx} T_n^{\rm FL}(\rho) = \frac{N P_{\rm tx}\sfl}{\rho B \log_2\left(1+\frac{g_n N P_{\rm tx}^{\rm FL}}{\rho B N_0}\right)}.
\end{equation}

\subsection{Throughput-oriented devices communication model}
For devices of set $\mathcal{K}$, we consider and compare two popular \ac{RA} protocols, namely, ALOHA and \ac{saloha}. The main difference is that \ac{saloha} assumes time is slotted, nodes have synchronized clocks, and a transmission can only start at the beginning of a slot. For analysis tractability, we consider that $K$ is big enough so that the infinite number of devices model is a good approximation. 

We assume that the transmission time $\tra$ is fixed and equal for every node, i.e., for any $k,\ell \in \mathcal{K}$ if the rate changes significantly due to different channel gains $g_k \neq g_\ell$, the nodes can adjust the size of the transmitted packet so that $\sra_k \neq \sra_\ell \implies {\left(\tra\right)}_k \approx {\left(\tra\right)}_\ell = \tra$. Hence, we will assume that, statistically, each node transmits a packet of $\sra$~bits with average channel gain $\bar{g}$.

Under this assumption, if we denote by $R^{\rm RA}(b)$ the Shannon rate for the considered \ac{RA} protocols with available bandwidth $b$, we have
\begin{equation}
    \label{eq:ra-rate}
    R^{\rm RA}(b) = b \log_2\left(1+\frac{\bar{g}P_{\rm tx}}{b N_0}\right),
\end{equation}
and the average \ac{RA} packet transmission time using a bandwidth $b$ is
\begin{equation}
    \label{eq:ra-tx-time}
    \begin{aligned}
    \tra &= \tra(B) {\rm P}[b=B] + \\ &\quad + \tra((1-\rho)B) {\rm P}[b=(1-\rho)B]=\\
         &= \frac{\left(T_{\rm tot}^{\rm FL}-\tfl\right) \sra}{T_{\rm tot}^{\rm FL} R^{\rm RA}(B)}+\frac{\tfl \sra}{T_{\rm tot}^{\rm FL} R^{\rm RA}((1-\rho)B)}.
    \end{aligned}
\end{equation}
For this formula, we considered that, for the share of time when \ac{FL} devices are not transmitting (either computing or staying idle) $\left(T_{\rm tot}^{\rm FL}-\tfl\right)/T_{\rm tot}^{\rm FL}$, the available bandwidth for the \ac{RA} devices is $B$. In contrast, during \ac{FL} model update $\tfl/T_{\rm tot}^{\rm FL}$, the available bandwidth becomes $(1-\rho)B$.

To measure the success probability of transmitting a packet without collisions, we assume that the packet arrivals follow a Poisson distribution $\mathcal{P}(\lambda)$. The success probability $\rm P_s$ is thus
\begin{equation}
\begin{aligned}
    \label{eq:success-probability}
    \rm P_s&= \rm P\left[\text{no coll.} \mid t \in \left[t_0,t_2\right]\right]\rm P\left[t \in \left[t_0,t_2\right]\right] + \\ &\quad +\rm P\left[\text{no coll.} \mid t \in \left[t_2,t_3\right]\right]\rm P\left[t \in \left[t_2,t_3\right]\right].
\end{aligned}
\end{equation}

\subsubsection{Pure ALOHA analysis}
For pure ALOHA, the vulnerable window is $W=2T_p$, where $T_p$ is the transmission duration. This translates into
\begin{equation}
    \label{eq:aloha-prob}
    \rm P_s^{\rm A} = \frac{T_{\rm tot}^{\rm FL}-\tfl}{T_{\rm tot}^{\rm FL}} e ^{-\lambda 2 \tra (B)} + \frac{\tfl}{T_{\rm tot}^{\rm FL}} e ^{-\lambda 2 \tra ((1-\rho)B)},
\end{equation}
with an expected normalized throughput given by
\begin{equation}
\begin{aligned}
    \label{eq:aloha-throughput}
    \rm Q^{\rm A}(\lambda, \rho) &= \lambda \tra (B) \frac{T_{\rm tot}^{\rm FL}-\tfl}{T_{\rm tot}^{\rm FL}} e ^{-\lambda 2 \tra (B)} + \\
    &\quad + \lambda \tra ((1-\rho)B)\frac{\tfl}{T_{\rm tot}^{\rm FL}} e ^{-\lambda 2 \tra ((1-\rho)B)},
\end{aligned}
\end{equation}
where $\lambda = \lambda^\prime+\mu$ is the total arrival rate composed of exogenous new arrivals $\lambda^\prime$ and retransmissions $\mu$. The average transmission energy, including retransmissions, is
\begin{equation}
    \label{eq:aloha-energy}
    E^{\rm A}(\lambda, \rho) = \frac{P_{\rm tx}^{\rm RA} \tra(\rho)}{\rm P_s^{\rm A}(\lambda, \rho)}. 
\end{equation}

\subsubsection{S-ALOHA analysis}
In \ac{saloha}, the vulnerable window is the slot duration $W=T_p$. Since there is no coordination among \ac{RA} and \ac{FL} devices, i.e., the throughput-oriented devices do not know when \ac{FL} devices access the medium, the worst case is taken as the vulnerable window. This amounts to considering $W=T_p=\tra((1-\rho)B)$, and, as a consequence, the success probability is
\begin{equation}
    \label{eq:saloha-prob}
    \rm P_s^{\rm SA} = e^{-\lambda \tra ((1-\rho)B)}.
\end{equation}
The expected normalized throughput is hence given by
\begin{equation}
    \label{eq:saloha-throughput}
    \rm Q^{\rm SA}(\lambda, \rho) = \lambda \tra e^{-\lambda \tra ((1-\rho)B)}
\end{equation}
and the average transmission energy, including retransmissions, is
\begin{equation}
    \label{eq:saloha-energy}
    E^{\rm SA}(\lambda, \rho) = \frac{P_{\rm tx}^{\rm RA} \tra(\rho)}{\rm P_s^{\rm SA}(\lambda, \rho)}. 
\end{equation}

%% file: sections/formulation.tex
\section{Joint Radio Resource Allocation}
We aim to minimize the global communication energy consumption subject to (i) a latency budget for the federated learning round and (ii) a minimum guaranteed throughput for RA traffic, by adjusting the bandwidth allocated to each set of devices. We also ask that a minimum retransmission rate be guaranteed for RA traffic for system stability.

\subsection{Problem formulation}
Using the system model introduced in Section~\ref{sec:system}, we formulate the following optimization problem controlling the bandwidth share $\rho$ and the total (re)transmission rate $\lambda$, for $a\in\{\rm A, \rm SA\}$, depending on the specific \ac{MAC} protocol used by random access devices.
\begin{equation}
    \label{eq:optimization-problem}
    \begin{aligned}
        \min_{\lambda,\rho} \quad & \left(\lambda^\prime \,T_{\rm tot}^{\rm FL}\right) E^{a}(\lambda,\rho) + \sum_{n\in\mathcal{N}} E_n^{\rm FL}(\rho)  \\
        \text{s.t.} \quad & {\rm Q}^a(\lambda, \rho) \ge q\\
                          & \tfl \le T_{\rm tot}^{\rm FL} - \tcpu \\
                          & 0 \le \rho \le 1\\
                          & \lambda - \lambda^\prime \ge \varepsilon
    \end{aligned}
\end{equation}
Here, we considered that the total number of newly generated attempts by the throughput-oriented devices is $(\lambda^\prime \,T_{\rm tot}^{\rm FL})$ (note that the energy expressions in Eqs.~\eqref{eq:aloha-energy} and~\eqref{eq:saloha-energy} already account for possible retransmissions).

The formulated optimization problem amounts to minimizing the overall energy consumption subject to (i) a minimum throughput guarantee $q$, (ii) a latency constraint for \ac{FL} model upload, (iii) the physical bounds for the bandwidth share allocation $\rho\in[0,1]$, and (iv) the request that the retransmission rate is at least $\varepsilon$ for throughput-oriented devices.

\subsection{Solution method}
\label{sec:solution}
The optimization problem~\eqref{eq:optimization-problem} is non-convex. Therefore, we adopt a decomposition approach, looking first for the optimal value $\lambda^\star(\rho)$. We observe that the throughput function ${\rm Q}^a(\lambda, \rho)$ is monotonic in $\lambda$. For any value of $\rho$, the feasible region is $\lambda \ge \lambda_{\min}(\rho)$, with ${\rm Q}^a(\lambda_{\min}, \rho) = q$. The throughput that minimizes the energy consumption is the one at the boundary of the constraint: choosing a higher $\lambda$ would imply having a higher retransmission rate, increasing the energy consumption. The optimal solution, combined with the minimum retransmission rate constraint, is thus
\begin{equation}
    \lambda^\star(\rho) = \max\{\lambda_{\min}(\rho), \lambda^\prime+\varepsilon\}.
\end{equation}
At this point, we are left with the optimization problem
\begin{equation}
\begin{aligned}
    \min_\rho \quad & f(\rho) = E\left(\lambda^\star(\rho), \rho\right)\\
    \text{s.t.} \quad & \tfl \le T_{\rm tot}^{\rm FL} - \tcpu \\
                          & 0 \le \rho \le 1
\end{aligned}
\end{equation}
where $E(\lambda, \rho)$ is the overall energy consumption of the system (random access and federated learning nodes). For this, we adopt a grid search over the feasible region $0 \le \rho \le 1$ and evaluate the objective function $f(\rho)$. The couple
\begin{equation}
    (\lambda^\star,\rho^\star) = \argmin_\rho E\left(\lambda^\star(\rho), \rho\right)
\end{equation}
is selected as the optimal solution to problem~\eqref{eq:optimization-problem}. The solution found is globally optimal, up to a resolution error due to the discretization interval.

%% file: sections/results.tex
\section{Numerical Results}
\label{sec:results}
In this section, we present the numerical results obtained for the optimization, performing
\begin{inparaenum}[(i)]
    \item an analytical/numerical study on the theoretical maximum throughput reachable by the \ac{RA} protocols, jointly with the related system energy consumption;
    \item a sensitivity study with respect to the number of \ac{FL} devices $N$ and the fresh packets arrival rate $\lambda^\prime$;
    \item a study on $5$ setup configurations, with details on the \ac{FL} model accuracy and learning duration, and energy- and throughput-optimized metrics.
\end{inparaenum}
For the simulations, unless otherwise specified, the parameters summarized in Tab.~\ref{tab:parameters} are used.

\begin{table}[tb]
\centering
\caption{Summary of parameters}
\label{tab:parameters}
\begin{tabular}{@{}llc@{}}
\toprule
\textbf{Description}  & \textbf{Parameter}   & \textbf{Value}       \\ \midrule
Number of FL devices & $N$                    & $30$                 \\
Rate of fresh RA arrivals & $\lambda^\prime$       & $10^4$~pkts/s   \\
Time reserved for FL computation & $\tcpu$                & $38$~s          \\
Total FL round duration & $T_{\rm tot}^{\rm FL}$ & $60$~s          \\
Minimum required throughput & $q$                    & $0.178$              \\
Minimum retransmissions rate & $\varepsilon$                    & $10^2$~pkts/s              \\
Total available bandwidth & $B$                    & $60$~MHz        \\
FL model size & $S^{\rm FL}$           & $100$~Mbits     \\
RA packet size & $S^{\rm RA}$           & $1.5$~kbits     \\
Channel gain & $\bar{g} = g_n$        & $0.1$                \\
Transmission power & $P_{\rm tx}^{\rm FL}=P_{\rm tx}^{\rm RA}$           & $0.4$~W         \\
Noise power spectral density & $N_0$                  & $10^{-17}$~W/Hz\\
\bottomrule
\end{tabular}%
\vspace{-0.3cm}
\end{table}

\subsection{Maximum throughput and energy consumption}

\begin{figure}[tb]
    \centering
    \includegraphics[width=\columnwidth]{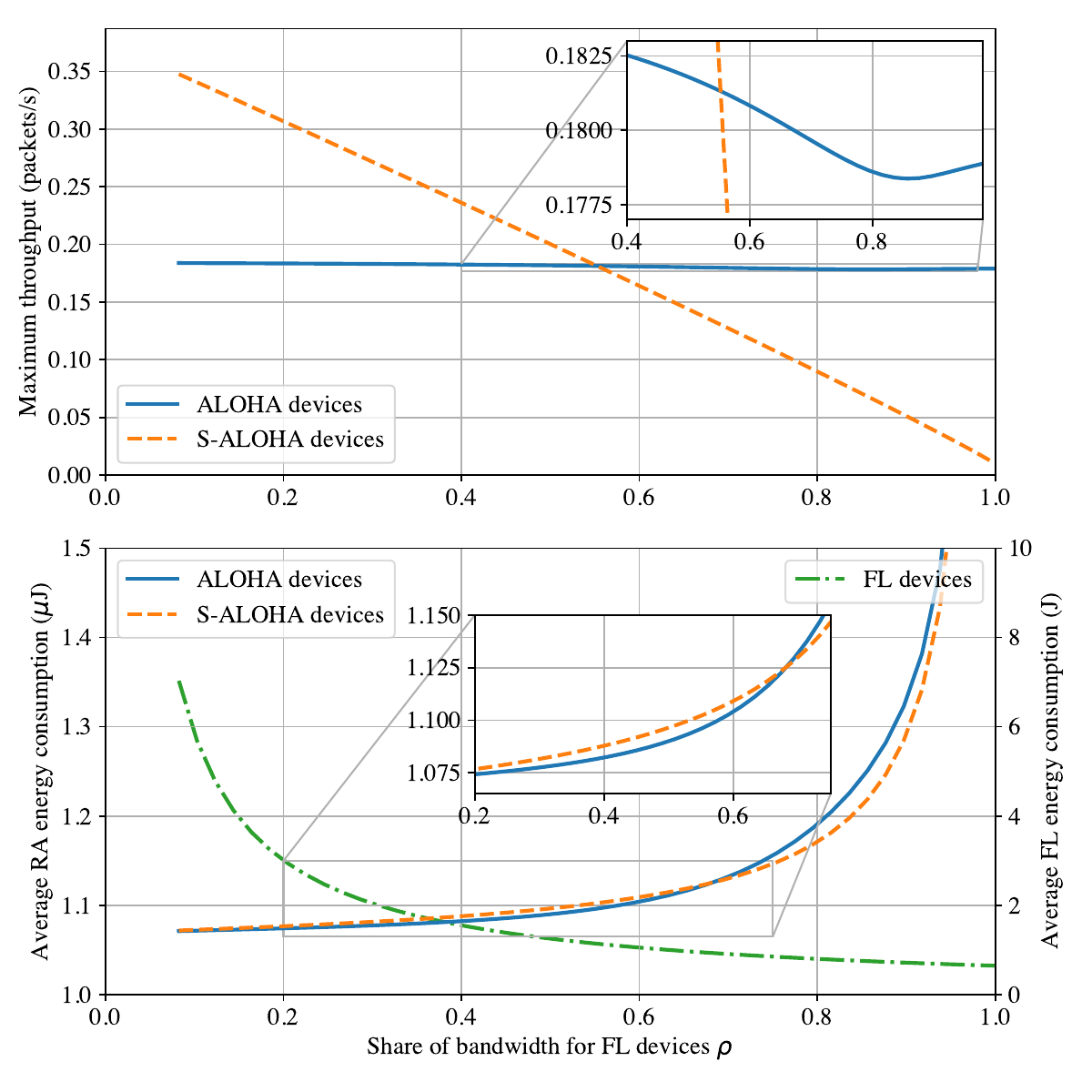}
    \caption{Top: maximum achievable throughput of the RA protocols as a function of the bandwidth share $\rho$. Bottom: FL and RA average energy consumption as a function of the bandwidth share $\rho$. Results obtained with $N=30$ and optimized $\lambda$.}
    \label{fig:throughput_energy}
    \vspace{-0.1cm}
\end{figure}

We analyze the theoretical maximum throughput reachable by \ac{RA} devices, shown in Fig.~\ref{fig:throughput_energy}, at the top. We observe that for $\rho < 0.1$, the problem is infeasible as FL devices cannot upload the updated model on time. From this threshold onward, the problem is feasible, and the theoretical throughput achievable by ALOHA and \ac{saloha} is initially very close to the maximum ($1/2e$ and $1/e$, respectively). However, while \ac{saloha}'s throughput linearly decreases with $\rho$, ALOHA's throughput is almost constant. It shows a minimum at $\rho\approx 0.82$ that can be seen in the zoomed area. This property is particularly desirable for ALOHA, as it enables it to achieve close-to-optimal throughput even when offering a high bandwidth to \ac{FL} devices. For $\rho < 0.55$, \ac{saloha} achieves higher throughput.

This result should be evaluated alongside the system energy consumption (Fig.~\ref{fig:throughput_energy}, bottom), which shows the average energy consumption for transmitting one packet (RA devices) and uploading one model (FL devices). Clearly, due to the different sizes of the updates, the consumption scale is different. FL energy consumption dominates the system's consumption unless the number of RA transmissions is at least $10^6$. For high $\rho$, \ac{saloha} can actually reduce the energy consumption, but this comes at the cost of a very low achievable throughput for RA devices. To keep throughput guarantees, \ac{saloha} needs to work with low $\rho$, increasing the energy consumption of FL devices. Hence, ALOHA offers higher flexibility and efficiency, unless $q > 1/2e$ is needed. In this case, \ac{saloha} is mandatory, at the cost of a very high FL energy cost.

\subsection{Sensitivity for FL devices and RA transmissions}

\begin{figure}[tb]
    \centering
    \includegraphics[width=\columnwidth]{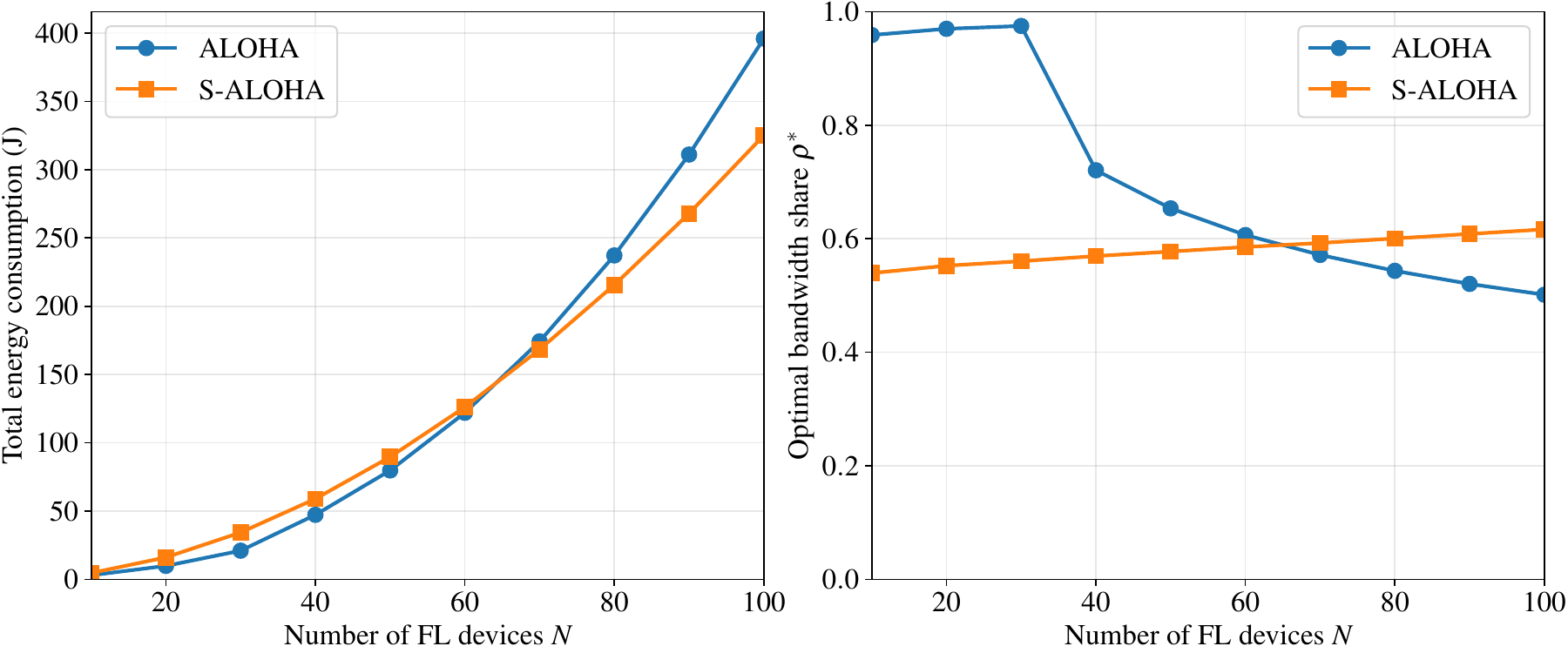}
    \caption{Optimized performance varying the number of FL devices $N$. Left: overall system energy consumption. Right: allocated bandwidth share to FL devices $\rho$. ($q=0.178$, $\lambda^\prime=10^4$ pkts/s).}
    \label{fig:sensitivity_FL}
    \vspace{-0.3cm}
\end{figure}

\begin{figure}[tb]
    \centering
    \includegraphics[width=\columnwidth]{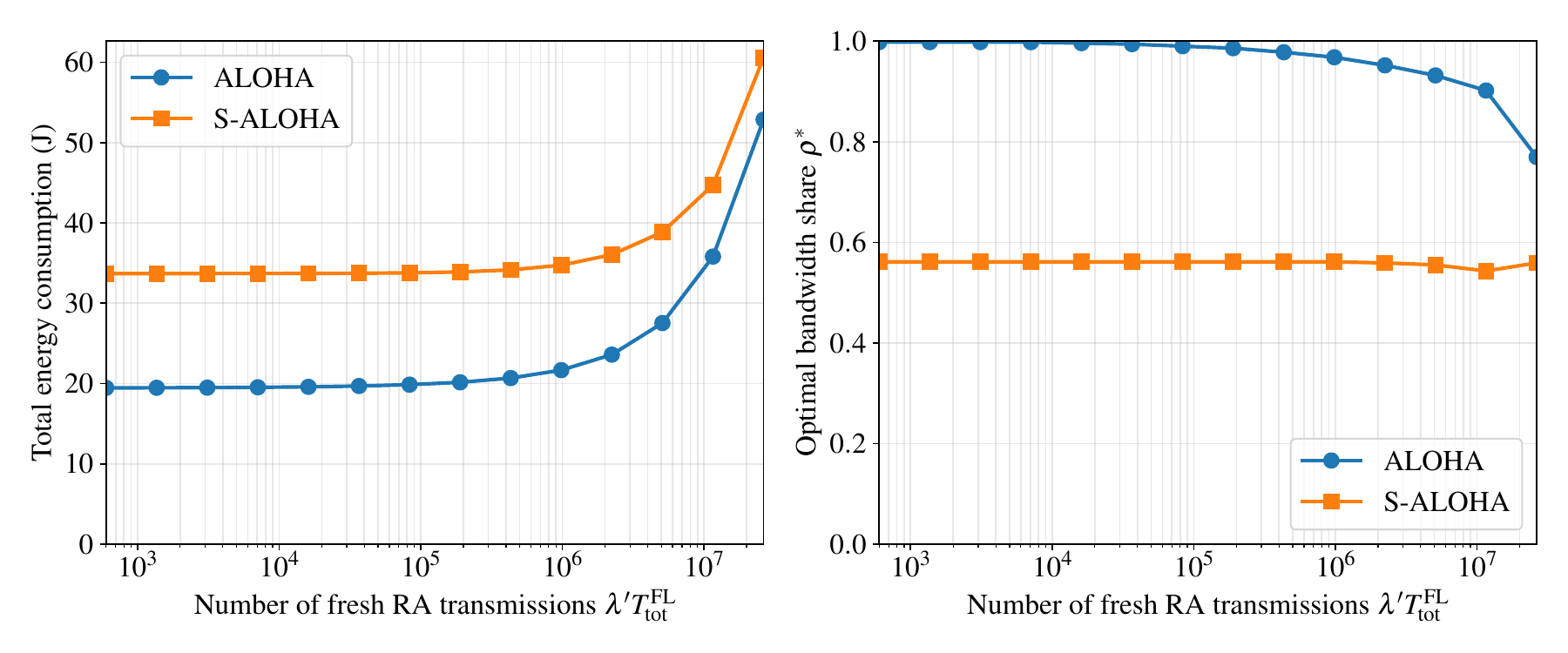}
    \caption{Optimized performance varying the number of RA fresh transmissions $\lambda^\prime T_{\rm tot}^{\rm FL}$ ($T_{\rm tot}^{\rm FL} = 60$~s). Left: overall system energy consumption. Right: allocated bandwidth share to FL devices $\rho$. ($q=0.178$, $N=30$).}
    \label{fig:sensitivity_RA}
    \vspace{-0.2cm}
\end{figure}

The procedure presented in Sect.~\ref{sec:solution} is used to present the optimization results in this section. Specifically, we show the capabilities of the system by varying the number of \ac{FL} devices $N$ (see Fig.~\ref{fig:sensitivity_FL}) and the number of RA fresh transmissions $\lambda^\prime T_{\rm tot}^{\rm cpu}$ (see Fig.~\ref{fig:sensitivity_RA}). 

As shown in Fig.~\ref{fig:sensitivity_FL}, with the chosen $q=0.178$, ALOHA can sustain the selection of a higher bandwidth share for FL devices until $N \approx 60$, threshold from which \ac{saloha} becomes more efficient. The energy consumption increases monotonically with the number of \ac{FL} devices when the RA transmissions are fixed, and, in general, the protocol supporting the highest $\rho$ is the most energy efficient, as, in this setting, FL consumption dominates. We note that the impact of the parameter $q$ is extremely relevant: allowing for a lower $q=0.17$ would make ALOHA always more efficient than \ac{saloha}, while increasing it to $0.182$ would make ALOHA sustainable only up to $N=20$.

In Fig.~\ref{fig:sensitivity_RA}, we show the performance varying the number of fresh arrivals of RA devices ($q=0.178$). \ac{saloha} is independent of $\lambda^\prime$, as the result is a constant line. On the other hand, the efficiency of ALOHA decreases as the number of transmissions increases. Here, the number of FL devices also affects performance, and the performance of ALOHA degrades more rapidly as $N$ grows, eventually becoming worse than \ac{saloha} for $N > 60$.

\subsection{Showcase of joint FL and RA optimization}

We test $5$ example configurations, summarized in Tab.~\ref{tab:configurations}, to showcase the potential of optimizing the network and highlight what \ac{RA} protocol should be chosen depending on the context.

\begin{table}[t]
\centering
\caption{Simulation configurations}
\label{tab:configurations}
\begin{tabular}{lccc}
\toprule
 & $N$ & $T_{\rm tot}^{\rm FL}$ [s] & $\lambda^\prime$ [pkts/s] \\
\midrule
$\mathcal{C}_1$ & 10 & 45  & $10^{4}$ \\
$\mathcal{C}_2$ & 10 & 45  & $10^{5}$ \\
$\mathcal{C}_3$ & 30 & 40  & $10^{5}$ \\
$\mathcal{C}_4$ & 30 & 60  & $10^{5}$ \\
$\mathcal{C}_5$ & 30 & 120 & $5 \times 10^{5}$ \\
\bottomrule
\end{tabular}
\vspace{-0.3cm}
\end{table}

\begin{figure}[t]
    \centering
    \includegraphics[width=\columnwidth]{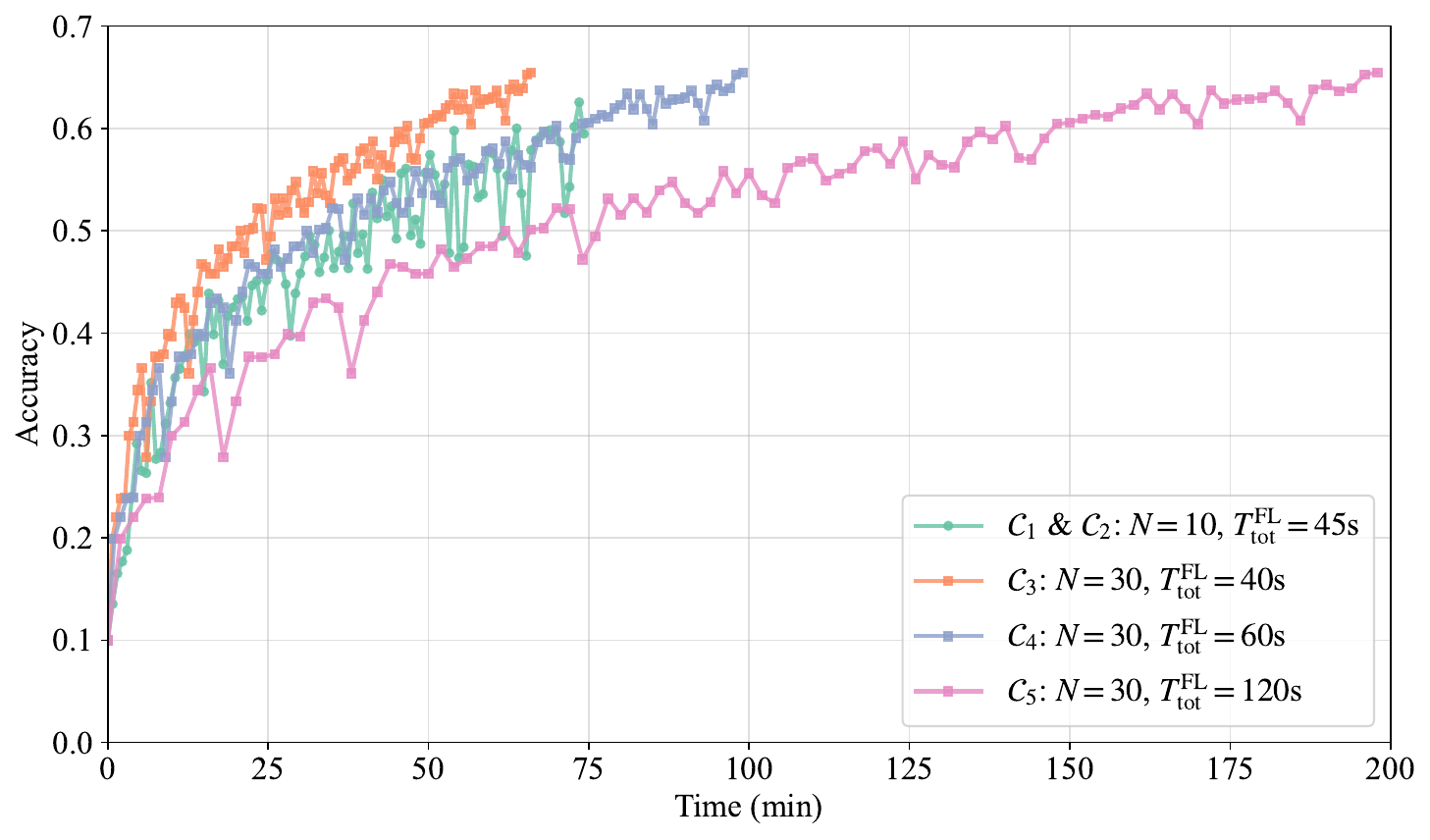}
    \caption{FL test accuracy for the five different configurations with respect to the time elapsed (every round lasts $T_{\rm tot}^{\rm FL}$~s, with $N$ randomly sampled clients for training). The number of clients randomly sampled for testing is always $30$, and they are sampled independently of the training set.}
    \label{fig:federated}
    \vspace{-0.3cm}
\end{figure}

The \ac{FL} results are obtained using the \texttt{flower} framework\footnote{\url{https://flower.ai}} and the popular \texttt{cifar10} dataset. We randomly sample $N$ clients from a pool of $M=100$ clients at each of the $C=100$ communication rounds, according to the FedAvg~\cite{mcmahan2017communication} aggregation strategy. The evaluation is always performed on $V=30$ clients, randomly sampled independently of the set~$\mathcal{N}$ and its cardinality. A simple convolutional neural network (CNN) with three convolutional layers and a linear layer before the softmax output is trained with Adam as a local solver. The size of the model, used as optimization parameter, was $\sfl = 36.72$~Mbits.
In Fig.~\ref{fig:federated} we show the \ac{FL} model accuracy over time for the five configurations, and the respective optimized metrics are reported in Tab.~\ref{tab:opt_summary}. 

\begin{table*}[t]
\centering
\caption{Summary of joint FL and RA optimization results for ALOHA (A) and S-ALOHA (SA).}
\label{tab:opt_summary}
\resizebox{\textwidth}{!}{%
\begin{tabular}{lcccccccccccccccccccc}
\toprule
 & \multicolumn{2}{c}{$\tfl~[\mathrm{s}]$}
 & \multicolumn{2}{c}{$\tra~[\mu\mathrm{s}]$}
 & \multicolumn{2}{c}{$E_b^{\mathrm{FL}}~[\mathrm{nJ/bit}]$}
 & \multicolumn{2}{c}{$E_b^{\mathrm{RA}}~[\mathrm{nJ/bit}]$}
 & \multicolumn{2}{c}{$E^{\mathrm{FL}}~[\mathrm{J}]$}
 & \multicolumn{2}{c}{$E^{\mathrm{RA}}~[\mathrm{J}]$}
 & \multicolumn{2}{c}{$E_{\mathrm{tot}}~[\mathrm{J}]$}
 & \multicolumn{2}{c}{$\rm P_{\mathrm{s}}$}
 & \multicolumn{2}{c}{$\rho^\star$}
 & \multicolumn{2}{c}{$\lambda^\star~[10^{5}\,\mathrm{pkts/s}]$} \\
\cmidrule(lr){2-3}\cmidrule(lr){4-5}\cmidrule(lr){6-7}\cmidrule(lr){8-9}
\cmidrule(lr){10-11}\cmidrule(lr){12-13}\cmidrule(lr){14-15}\cmidrule(lr){16-17}
\cmidrule(lr){18-19}\cmidrule(lr){20-21}
 & A & SA & A & SA & A & SA & A & SA & A & SA & A & SA & A & SA & A & SA & A & SA & A & SA \\
\midrule
$\mathcal{C}_1$ & 0.22 & 0.38 & 1.05 & 0.97 & 2.37 & 4.19 & 0.62 & 0.57 & 0.87 & 1.54 & 0.42 & 0.38 & \textbf{1.29} & 1.92 & 0.46 & 0.46 & 0.96 & 0.53 & 4.06 & 4.02 \\
$\mathcal{C}_2$ & 0.24 & 0.47 & 0.99 & 0.97 & 2.58 & 5.16 & 0.58 & 0.42 & 0.95 & 1.90 & 3.93 & 2.82 & 4.88 & \textbf{4.71} & 0.46 & 0.62 & 0.88 & 0.42 & 4.07 & 2.97 \\
$\mathcal{C}_3$ & 0.64 & 1.08 & 1.13 & 0.99 & 6.96 & 11.76 & 0.70 & 0.57 & 7.67 & 12.96 & 4.21 & 3.43 & \textbf{11.87} & 16.38 & 0.43 & 0.46 & 0.93 & 0.53 & 4.30 & 3.90 \\
$\mathcal{C}_4$ & 0.64 & 1.11 & 1.07 & 0.98 & 6.95 & 12.05 & 0.65 & 0.54 & 7.66 & 13.28 & 5.83 & 4.84 & \textbf{13.49} & 18.11 & 0.44 & 0.49 & 0.93 & 0.52 & 4.18 & 3.74 \\
$\mathcal{C}_5$ & 0.70 & 1.60 & 0.99 & 0.97 & 7.66 & 17.43 & 0.69 & 0.53 & 8.44 & 19.20 & 62.41 & 48.10 & 70.85 & \textbf{67.30} & 0.38 & 0.48 & 0.84 & 0.35 & 5.00 & 5.00 \\
\bottomrule
\end{tabular}%
}
\end{table*}

The two configurations $\mathcal{C}_1$ and $\mathcal{C}_2$ show identical \ac{FL} accuracy and overall round duration, as they are tested for the same number of FL devices $N=10$ and total available time $T_{\rm tot}^{\rm FL}=45$~s. However, they differ in terms of RA traffic rate: the RA traffic of $\mathcal C_2$ is ten times higher. Due to this, the allocated bandwidth share $\rho^\star$ drops for both ALOHA (A) and \ac{saloha} (SA). In both cases, A allocates a higher bandwidth to FL devices than SA. However, the energy consumption is dominated by FL devices in $\mathcal C_1$ and by RA devices in $\mathcal C_2$. ALOHA is $48.8\%$ more energy-efficient for $\mathcal C_1$, while \ac{saloha} is $3.6\%$ more energy-efficient in $\mathcal C_2$. For configurations $\mathcal C_3-\mathcal C_5$, with $N=30$, we progressively increase the round duration. As shown in Fig.~\ref{fig:federated}, the model accuracy is higher when sampling $30$ clients rather than $10$ ($6\%$ higher). The energy consumption, however, differs significantly for the different setups. To evaluate the consumption fairly regardless of the total experiment duration, we also report the per-bit energy consumption $E_b^{\rm FL}$ and $E_b^{\rm RA}$, for FL and RA, respectively. As reported in Tab.~\ref{tab:opt_summary}, it can be observed that $E_b^{\rm FL}$ for $\mathcal C_3 - \mathcal C_5$ is about three times the one of $\mathcal C_1 - \mathcal C_2$. This is caused by the fact that triplicating the number of FL devices produces a $\times 3$ increase in the transmission time $\tfl$ if the bandwidth share is not increased. However, increasing the bandwidth share allocated to FL while lowering energy consumption would cause the throughput to fall below the required threshold $q=0.178$. For configurations $\mathcal C_3$ and $\mathcal C_4$, A is more energy-efficient than SA by $38\%$ and $34.2\%$, respectively. On the other hand, SA is more energy-efficient by $5.3\%$ in $\mathcal C_5$.

Overall, by exploring the complex landscape of the joint optimization of FL devices (FDMA radio access) and RA throughput-oriented devices (ALOHA vs.\ \ac{saloha} protocols), we can infer the following two considerations:
\begin{itemize}
    \item ALOHA is more energy-efficient when the system's energy consumption is dominated by FL devices $\left(N\sfl \gg \lambda^\prime T_{\rm tot}^{\rm FL}\sra\right)$, as it allows for higher bandwidth shares $\rho^\star$ and hence lower $\tfl$.
    \item While the RA arrival rate $\lambda^\prime$ and/or the total round duration $T_{\rm tot}^{\rm FL}$ increase, RA devices' consumption starts to dominate the overall consumption. In these cases, \ac{saloha} can support higher throughputs and hence lower consumption: see the higher success probability $\rm P_{\rm s}$ for $\mathcal C_2$ and $\mathcal C_5$ in Tab.~\ref{tab:opt_summary}.
\end{itemize}

%% file: sections/concl.tex
\section{Conclusion}
In this paper, we study concurrent uplink transmissions from two sets of devices: \ac{FL} devices accessing the medium via \ac{FDMA}, and throughput-oriented devices using \ac{RA}, for which we compare the ALOHA and \ac{saloha} protocols. We formulated an optimization problem to minimize system energy consumption subject to \ac{FL} maximum round duration and a minimum guaranteed \ac{RA} throughput, with the bandwidth allocation share ($\rho$) and the (re)transmission rate ($\lambda$) as decision variables. Since this is a non-convex problem, the solutions are derived via grid search. By exploring the complex landscape of the optimized solutions, we show that neither \ac{RA} protocol universally outperforms the other: ALOHA is preferable when the system is dominated by \ac{FL} consumption, whereas \ac{saloha} achieves potentially better performance when the system is instead dominated by \ac{RA} traffic.